\documentclass[aps,prl,longbibliography,showpacs,twocolumn,superscriptaddress]{revtex4-2}
\usepackage{amsmath,amssymb,amsfonts,bm}
\usepackage{booktabs}
\usepackage{braket}
\usepackage{graphicx}
\usepackage{epstopdf}
\usepackage{dcolumn}
\usepackage{mathrsfs}
\usepackage{bbold}
\usepackage{dsfont}
\usepackage{float}
\usepackage{soul}
\usepackage[colorlinks=true,linkcolor=blue,citecolor=blue, urlcolor=blue,bookmarks=false]{hyperref}

\usepackage{tgtermes}
\usepackage{multirow}

\begin{document}
	
\title{Theory of $d+id$ Second-Order Topological Superconductors}
\author{Zi-Ming Wang}
\thanks{These authors contributed equally to this work.}
\affiliation{Department of Physics and Chongqing Key Laboratory for Strongly Coupled Physics, Chongqing University, Chongqing 400044,  China}

\author{Meng Zeng}
\thanks{These authors contributed equally to this work.}
\affiliation{Department of Physics, University of California, San Diego, California 92093, USA}

\author{Chen Lu}
\thanks{These authors contributed equally to this work.}
\affiliation{New Cornerstone Science Laboratory, Department of Physics, School of Science, Westlake University, Hangzhou 310024, Zhejiang, China}

\author{Da-Shuai Ma}
\affiliation{Department of Physics and Chongqing Key Laboratory for Strongly Coupled Physics, Chongqing University, Chongqing 400044, China}
\author{Rui-Xing Zhang}
\affiliation{Department of Physics and Astronomy, University of Tennessee, Knoxville, Tennessee 37996, USA}
\affiliation{Department of Materials Science and Engineering, University of Tennessee, Knoxville, Tennessee 37996, USA}
	
\author{Lun-Hui Hu}
\email{hu.lunhui.zju@gmail.com}
\affiliation{Department of Physics and Astronomy, University of Tennessee, Knoxville, Tennessee 37996, USA}
\affiliation{Center for Correlated Matter and School of Physics, Zhejiang University, Hangzhou 310058, China}
\affiliation{Department of Applied Physics, Aalto University School of Science, FI-00076 Aalto, Finland}

\author{Dong-Hui Xu}
\email{donghuixu@cqu.edu.cn}
\affiliation{Department of Physics and Chongqing Key Laboratory for Strongly Coupled Physics, Chongqing University, Chongqing 400044, China}

\date{\today}
	
\begin{abstract}
Topological superconductors are a class of unconventional superconducting materials featuring sub-gap zero-energy Majorana bound modes that hold promise as a building block for topological quantum computing. In this work, we study the realization of second-order topology that defines anomalous gapless boundary modes in a two-orbital superconductor with spin-orbital couplings. We reveal a time-reversal symmetry-breaking second-order topological superconducting phase with $d+id$-wave orbital-dependent paring without the need for the external magnetic field. Remarkably, this orbital-active $d$-wave paring gives rise to anomalous zero-energy Majorana corner modes, which is in contrast to conventional chiral $d$-wave pairing, accommodating one-dimensional Majorana edge modes. Our work not only reveals a unique mechanism of time-reversal symmetry breaking second-order topological superconductors but also bridges the gap between second-order topology and orbital-dependent pairings. 
\end{abstract}
	
\maketitle
	
\emph{Introduction.}---
Topological superconductors~(TSCs) are exotic quantum condensed phases of matter with topologically nontrivial structures of Cooper pair wavefunctions. As perhaps the most remarkable consequence of TSCs, spatially localized Majorana zero modes (MZMs) can be trapped in vortex cores of a two-dimensional~(2D) $p$-wave TSC~\cite{Read:00,Volovik:99,Ivanov:01} or be formed at the ends of a one-dimensional $p$-wave superconductor~\cite{kitaev2001unpaired}. MZMs manifest non-Abelian quantum statistics, which naturally encode topological qubits that pave the way for fault-tolerant quantum computation~\cite{Kitaev:03,Nayak:08}. Although naturally occurring TSC materials appear rare and elusive, the past few decades have witnessed a tremendous effort to discover artificial topological superconductivity in various quantum materials~\cite{beenakker2011search,alicea2012new,Elliott2015colloquium,sato2017topological}, following the pioneering theories~\cite{fu2008superconducting,sato2009nonabelian,sau2010generic}. So far, evidence of MZMs has been experimentally reported in several systems, ranging from one-dimensional superconducting hybrids~\cite{lutchyn2018majorana,flensberg2021engineered,nadj2014observation} to vortex cores on a proximitized topological insulator surface~\cite{Sun2016majorana} or on an iron-based superconductor surface~\cite{wang2018evidence}. 

The recent advances of topological band theory have unveiled an entirely new category of ``higher-order" TSCs with an unprecedented bulk-boundary relation~\cite{Langbehn2017PRL,YanPRL2018,Wang2018PRL,Zhuprb2018,khalaf2018higher,geier2018second,Yuxuan2018PRB,Tao2018PRB,hsu2018majorana,shapourian2018topological,Volpezprl2019,Liu2019PRL,Fulga2019PRB,Yan2019PRL,ZengPRL19,zhu2019second,Bultinck2019PRB,ZhangRX19PRL1,ZhangRX19PRL2,wu2019higher,ghorashi2019second,yan2019majorana,peng2019proximity,you2019higher,hsu2020inversion,wu2020boundary,Kheirkhah2020first,Kheirkhah2020majorana,Wuprl2020corner,ahn2020higher,zhang2020topological,zhang2020detection,roy2020higher,ghorashi2020vortex,pahomi2020braiding,tiwari2020chiral,Schindler2020pairing,roy2021mixed,Varjas2019PRL,chenruiprl,li2021higerorder,fu2021chiral,Ikegaya2021tunable,qin2022topological,Scammell2022intrinsic}. For example, in two dimensions (2D), a second-order TSC generally binds 0D MZMs around the geometric corners of a finite-size system, which is in contrast with a ``conventional" TSC. In pursuit of corner MZMs, a crucial conceptual question to look for new simple and feasible recipes that are applicable to real-world superconductors. Given the important role of orbital degrees of freedom in unconventional superconducting systems, a comprehension of whether multi-orbiral pairing can enable higher-order TSC is certainly necessary but still largely incomplete~\cite{Ong2013tetrahedral,ong2016concealed,nica2017orbital,Zeng2023spin-orbital}.

In this work, we show that orbital-active pairing can stabilize second-order class-D TSC that is protected by a $C_4$ rotation symmetry. The topological nature of the TSC phase is confirmed by numerically revealing the Majorana corner modes, as well as a topological quantum chemistry analysis. Besides sample corners, lattice dislocations are also found to trap MZMs as a result of the inherent weak topology, which makes our system a viable platform for designing and building Majorana qubits. 
 
 
\emph{ Model of $d+id$ second-order TSCs and symmetry analysis.}---We consider a normal state two-orbital  $\{d_{xz},d_{yz}\}$ tight-binding model on the square lattice with both asymmetric SOC (i.e.~Rashba type) and on-site SOC,
\begin{align}
H_\text{n}=&\sum_{\bm k}\Psi^{\dagger}\left(\bm{k}\right)\{ \epsilon \left(\bm{k}\right) \sigma_{0} s_0 +\widetilde{\epsilon} \left(\bm{k}\right) \sigma_{z} s_0 +\epsilon^{\prime \prime} \left(\bm{k}\right) \sigma_{x} s_0\nonumber\\
+&\lambda_\text{I}\sigma_{y}s_{z} 
+\lambda_\text{R} \sin k_x \sigma_{0} s_y -\lambda_\text{R} \sin k_y \sigma_{0} s_x\} \Psi\left(\bm{k}\right),
\label{Normal}
\end{align}
where $ \Psi^{\dagger}\left(\bm{k}\right)=({c}^{\dagger}_{d_{xz},\uparrow},{c}^{\dagger}_{d_{xz},\downarrow}, {c}^{\dagger}_{d_{yz},\uparrow},{c}^{\dagger}_{d_{yz},\downarrow}) $, $ \epsilon_{\textbf{k}}= -2t\cos k_x -2t\cos k_y +4t -\mu $, $ \widetilde{\epsilon} \left(\bm{k}\right) =-2\widetilde{t} \cos k_{x}+2\widetilde{t} \cos k_{y} $,  and $ \epsilon^{\prime \prime} \left(\bm{k}\right) =2 t^{\prime \prime} \sin k_{x} \sin k_{y} $. $ {\sigma}_{x,y,z} $ and $ {s}_{x,y,z} $ are the Pauli matrices for the orbital and spin degrees of freedom, $ \sigma_{0} $ and $ s_{0} $ are two $2 \times 2$ identity matrices. $ t $ describes the intra-orbital nearest-neighbor hopping, $ \widetilde{t} $ depicts the hopping anisotropy along the different direction of $ d_{xz} $, $ d_{yz} $ orbitals and $ t^{\prime \prime} $ is the inter-orbital next-nearest-neighbor hopping. The $ \lambda_\text{I} $ and $ \lambda_\text{R} $ are the strength of intrinsic and Rashba SOCs, respectively. The intrinsic SOC originates from the atomic spin-orbit coupling \cite{Vafek2017hund,Lee2010quasiparticle,Clepkens2021shadowed,boker2019quasiparticle}, while Rashba SOC can be intrinsic or externally induced by substrate effect. This normal Hamiltonian breaks inversion symmetry but preserves TRS: ${\mathcal{T}} {H_\text{n}}\left(\bm{k}\right) {\mathcal{T}}^{- 1}={H_\text{n}} \left(\bm{-k}\right)$, where $\mathcal{T}=i \sigma_0 s_y \mathcal{K}$ with $\mathcal{K}$ the complex conjugation operator. In addition, the normal Hamiltonian has the four-fold rotation symmetry ${r}_{4z} = i \sigma_y e^{-i\pi s_{z} /4} $.

To study the superconductivity, we define the Nambu basis $\left\{ \Psi^{\dagger}\left(\bm{k}\right),\Psi^{T}\left(\bm{-k}\right) \right\}$ and construct the Bogoliubov-de-Gennes (BdG) Hamiltonian as
\begin{equation}
H=\left[\begin{array}{cc}
H_{\text{n}}\left(\bm{k}\right) & \Delta\left(\bm{k}\right)\\
\Delta^{\dagger}\left(\bm{k}\right) & -H_{\text{n}}^{T}\left(\bm{-k}\right)
\end{array}\right].
\label{H_BdG}
\end{equation}
Here, the pairing potential $\Delta\left(\bm{k}\right)$ consists of both orbital-independent and orbital-dependent pairings, and can be generally written as,
\begin{align}	\Delta\left(\bm{k}\right)=\left[\Delta_\text{i}\Phi\left(\bm{k}\right)\sigma_{0} + \Delta_\text{o}(\boldsymbol{d}_\text{o}\left(\bm{k}\right) \cdot \boldsymbol{\sigma}) \right] i s_y.
\label{Pairing}
\end{align}
Here $\Delta_\text{i}$ and $\Delta_\text{o}$ are pairing amplitudes in orbital-independent and orbital-dependent channels, respectively. 
For our purpose, we are particularly interested in the d-wave pairing sector, thus the orbital-independent part is typically chosen as $\Phi\left(\bm{k}\right)=-\cos k_x + \cos k_y$. Without breaking the crystalline symmetry, a uniform orbital-dependent pairing $\boldsymbol{d}_\text{o}\left(\bm{k}\right)=(0,0,1)$ is also allowed.
For example, they belong to the same irreducible representation ($B_1$) of the $C_{4v}$ point group~\cite{nakayama2018nodal,nica2021multiorbital,smidman2023rmp}.
In the Supplementary Materials (SM), we self-consistently calculate the above gap function by using random phase approximation in the absence of SOCs and further find a spontaneous TRS breaking $d+id$ pairing by minimizing the free energy. Different from the traditional $d_{xy}+id_{x^2-y^2}$ (or $B_{1}+iB_{2}$), this $d+id$ pairing (or $B_{1}+iB_{1}$) preserves mirror symmetry, which enforces the vanishing of  the BdG Chern number for Eq.~\eqref{H_BdG}. While the first-order topology has thus been ruled out, we will show below that 2nd-order TSC can emerge naturally~\cite{bradlyn2017topological,Elcoro:ks5574,PhysRevE.96.023310,PhysRevB.97.035139,Po2017}.

\begin{figure}[!htbp]
\includegraphics[width=0.5\textwidth]{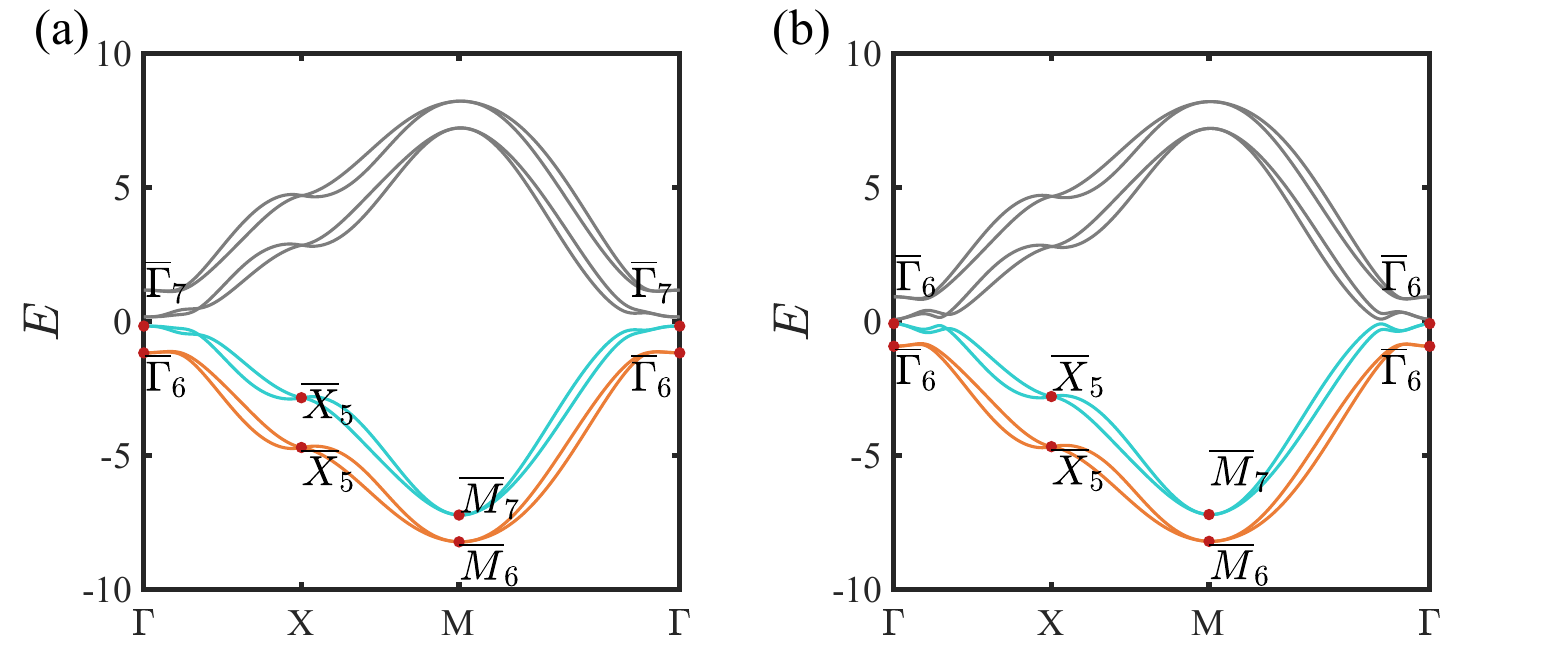}
\caption{ Bulk band structure and elementary band representations (EBRs) at high-symmetry points of the TRS breaking $d+id$ superconductor. (a) Topologically trivial superconducting phase for $\Delta_\text{o}=0.6$. (b) ${C}_4$ protected second-order TSC for $\Delta_\text{o}=0.3$. Using EBRs analysis from double space group $P4mm$, We can obtain ${C}_4$ index $J_z$ at $\Gamma$-point and M-point which is also $z$-direction angular momentum, where $J_z=\pm \frac{3}{2}$ of ${\Gamma}_{6}$ or ${M}_{6}$ and $J_z=\pm \frac{1}{2}$ of ${\Gamma}_{7}$ or ${M}_{7}$. Common parameters: $t=1$, $\widetilde{t}=0.2$, $t^{\prime \prime}=0.1$, $\mu=0.3$, $\lambda_{I}=0.5$, $\lambda_{R}=0.3$, $\Delta_\text{i}=0.1$.}
\label{Figure1}
\end{figure}

Because of the $d$-wave pairing $r_{4z} \Delta({\bf k}) r_{4z}^T = -\Delta(C_{4}^{-1}{\bf k})$, the BdG Hamiltonian preserves 
${{C}}_{4z}= r_{4z} \oplus -r_{4z}^\ast$, together with other symmetries $C_{2x}=i {\tau}_z {\sigma}_{0} s_z$ and $M_x=i {\tau}_z {\sigma}_{z} s_x$, we are capable of diagnosing the topology of the superconducting spectrum once it is fully gapped.
In Fig.~\ref{Figure1}, we present the band structures of superconductors with fully gapped trivial and higher-order topological superconductor phases.
In order to diagnose spatial symmetry-protected topological states, we employ the topological quantum chemistry theory to obtain symmetry-data vector $B$ \cite{PhysRevLett.120.266401,AroyoPerezMatoCapillasKroumovaIvantchevMadariagaKirovWondratschek+2006+15+27,Aroyo:xo5013,Aroyo2011183,10.1126/science.aaz7650} that is constituted by irreducible representations (irreps) of little groups at the maximal momenta in the first Brillouin Zone, as shown in TABLE~\ref{Table1} and inserted in Fig.~\ref{Figure1}. 
Referring to TABLE~\ref{Table1}, we can see that the topological trivial system, as shown in Fig.~\ref{Figure1}(a), is equivalent to that of a $s$ and $p_z$ orbitals at Wyckoff position of $1a$. 
In sharp contrast, the higher-order topological phase in Fig.~\ref{Figure1}(b), is equivalent to that of two $p_z$ orbitals at the  $1a$ and $1b$ Wyckoff positions. 
Notice that the $1b$ site is at the center of the square lattice and cannot be occupied by any orbitals in real space. Thus, the potential spatial symmetry-protected topological states fall within the scope of a superconducting analogue of obstructed atomic insulator (OAI), whose BdG Wannier orbitals are displaced from the lattice sites. The AOI can be effectively diagnosed by the real space invariant (RSI) defined at the $1b$ site. 
The RSI method determines the non-trivial second-order topology as $\left( \delta_1 , \delta_2 \right)=\left( -1 , \pm 1 \right)$, and the trivial phase is represented as $\left( \delta_1 , \delta_2 \right)=\left( 0 , 0 \right)$. The RSI $\left( \delta_1, \delta_2 \right)$ with SOC and broken TRS is defined in the SM.

\begin{table}[!htbp]
\caption{\label{Table1}
 The EBRs of space group $P4mm$ with SOC. The first row labels the Wyckoff positions. The second row is the irreducible representation of the double space group. The third row represents the orbits that induce the irreducible representation. The fourth row is the irreducible representations at high-symmetry points.}
\begin{ruledtabular}
\tabcolsep=0.1cm
\renewcommand{\arraystretch}{1.5}
\begin{tabular}{cccccccc}
 WPs  & \multicolumn{2}{c}{$1a$ $(000)$} & \multicolumn{2}{c}{$1b$ $(\frac{1}{2} \frac{1}{2} 0)$} \\

 EBRs  &  ${\overline{E}}_{1}\uparrow G(2)$  &  ${\overline{E}}_{2}\uparrow G(2)$  &  ${\overline{E}}_{1}\uparrow G(2)$  &  ${\overline{E}}_{2}\uparrow G(2)$  \\

 Orbitals  &  $s$  &  $p_z$  &  $s$ &  $p_z$  \\

\hline
 ${\overline{\Gamma}}$ $(000)$  &  ${\overline{\Gamma}}_{7}(2)$  &  ${\overline{\Gamma}}_{6}(2)$  &  ${\overline{\Gamma}}_{7}(2)$  &  ${\overline{\Gamma}}_{6}(2)$  \\
 ${\overline{M}}$ $(\frac{1}{2} \frac{1}{2} 0)$  &  ${\overline{M}}_{7}(2)$  &  ${\overline{M}}_{6}(2)$  &  ${\overline{M}}_{6}(2)$  &  ${\overline{M}}_{7}(2)$  \\
 ${\overline{X}}$ $(\frac{1}{2} 0 0)$  &  ${\overline{X}}_{5}(2)$  &  ${\overline{X}}_{5}(2)$  &  ${\overline{X}}_{5}(2)$  &  ${\overline{X}}_{5}(2)$  \\
\end{tabular}
\end{ruledtabular}
\end{table}

\begin{figure}[!htbp]
\includegraphics[width=0.5\textwidth]{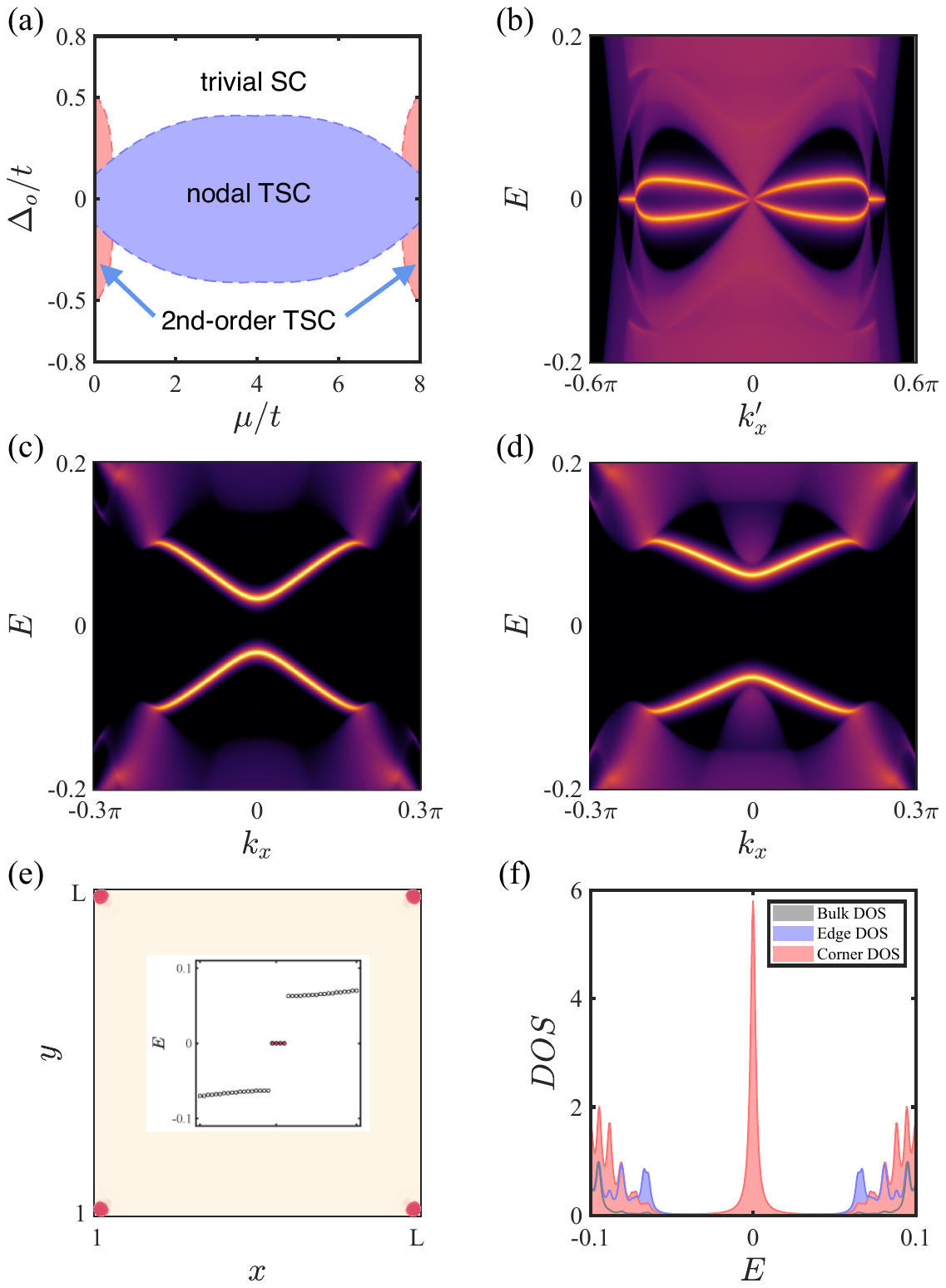}
\caption{ (a) Superconducting phase diagram on the plane of $\Delta_\text{o}$ and  $\mu$. The white region represents the trivial phase, the blue region is the nodal superconducting phase and the red region is the second-order TSC phase. (b) The spectral function of the nodal superconductor for open boundary condition along [1,1] when $\Delta_\text{o}=0.01$. There are two pairs of bulk nodes, and each pair of nodal points is connected by Majorana flat-band edge states. All the nodes are gapped out when $\Delta_\text{o} \neq 0$, as shown in (c-d). The energy gap increases as $\Delta_\text{i}$ is increased. (c) The spectral function for open boundary condition along [01] when $\Delta_\text{i}=0.05$, and (d) is $\Delta_\text{i}=0.1$. (e) The wave function profile of the four corner modes, where the inset shows the finite system spectrum near zero energy and four Majorana zero modes (red dots). The sample size is $120\times120$. (f) The density of states near zero energy with the density distributions of bulk, edge, and corner. Common parameters unless otherwise specified: $t=1$, $\widetilde{t}=0.2$, $t^{\prime \prime}=0.1$, $\mu=0.3$, $\lambda_{I}=0.5$, $\lambda_{R}=0.5$, $\Delta_\text{o}=0.3$, $\Delta_\text{i}=0.1$.}
\label{Figure2}
\end{figure}

\emph{Superconducting phase diagram.}---In Fig.~\ref{Figure2}(a), we present the superconducting phase diagram, which contains nodal superconductor, second-order TSC, and trivial phases, on the plane of $\mu$ and $\Delta_\text{o}$.
The gap closing and reopening of bulk dispersion at off-high-symmetry points (${\bf k}\notin\{\Gamma, X, Y, M\}$) distinguishes a nodal superconductor from a fully gapped one. In the fully gapped phase, we use the RSI method discussed above to determine its bulk topology. First, the gap function changes sign with respect to the reflection line along the [11] or [$1\bar{1}$] directions, suggesting a mirror symmetry-protected nodal superconductor~\cite{chiu2014classification}, which is highlighted in blue in Fig.~\ref{Figure2} (a). For example, it must be a nodal $d$-wave superconductor in the limit $\Delta_\text{o}=0$. More details can be found in the SM. 
Furthermore, the fully gapped superconductor can be either topologically trivial or nontrivial as we discussed above.  
When $\Delta_\text{o}$ is large enough, it is a fully gapped but trivial phase (the white region in Fig.~\ref{Figure2} (a)). The red region represents the second-order TSC phase. On the other hand, the TRS-breaking nodal superconductor is also topological, whose bulk nodes are protected by the mirror symmetry. Namely, the topological nodes are stable along the mirror-invariant lines (i.e.,~movable but irremovable by local perturbations). To show that, we perform a slab calculation with open boundary condition along the [$1\bar{1}$] direction and find the Majorana flat band states connecting two bulk nodes, as shown in $E_{k_x'}$ of Fig.~\ref{Figure2} (b). At fixed $k_x'$, the 1D Hamiltonian $H(k_y')$ has its own particle-hole symmetry (mirror$\otimes$PHS), leading to the $\mathbb{Z}_2$ topology invariant, as discussed in the SM.

We next consider the second-order TSC phase when $\mu$ is ``approaching'' the band bottom or top. First, we use Green's function method to calculate the spectral function along the [10] or [01] direction, which shows fully gapped features in Figs.~\ref{Figure2} (c) and (d), where the gap opening for the in-gap edge states is increased by increasing $\Delta_\text{i}$ (the gap of edge states $\sim 0.06$ in (c) and $\sim 0.12$ in (d)).  This will be explained later by constructing a low-energy edge theory. Then, we take the full tight-binding simulation on a square lattice to show the Majorana corner states in Fig.~\ref{Figure2} (e), where the inset shows the energy spectra with four zero energy states. We further calculate the local density of states (DOSs) in Fig.~\ref{Figure2} (f) to detect the Majorana zero modes. As we expect, both bulk and edge DOSs show a ``U'' shape, while a share zero bias peak is found for the DOS measured at corners.
\begin{figure*}[!htbp]
\includegraphics[width=\textwidth]{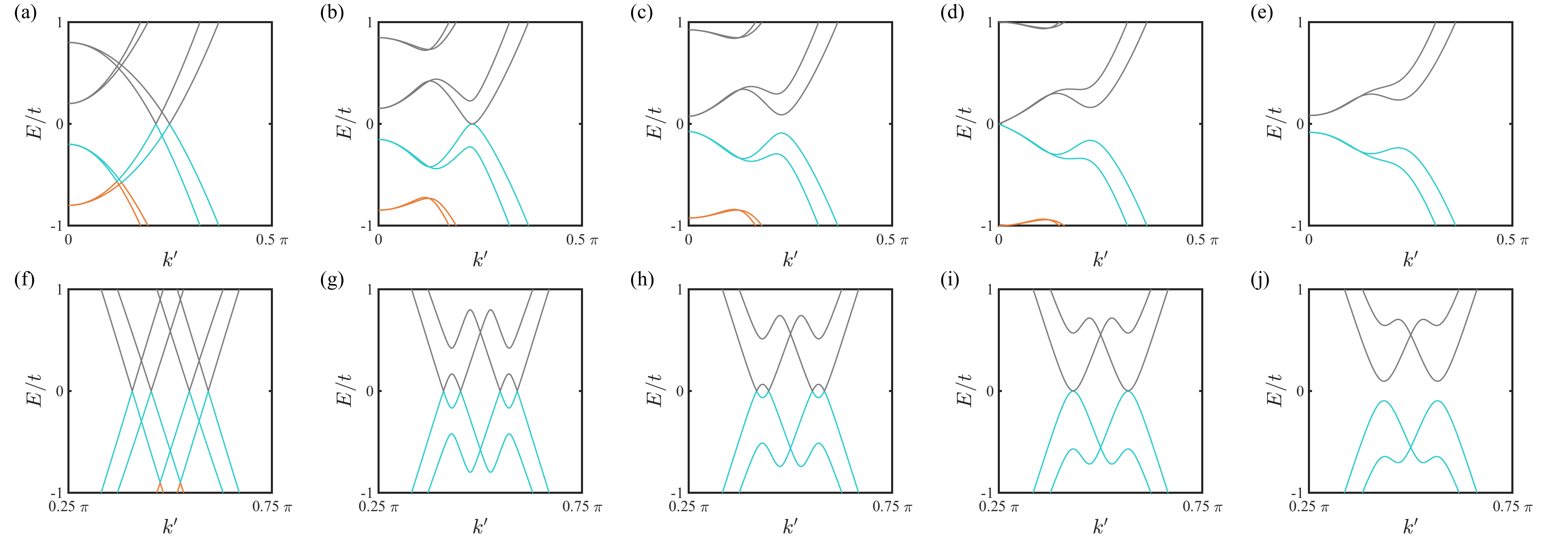}
\caption{
Bulk energy spectrum evolution under increasing $\Delta_\text{o}$ from $\Delta_\text{o}=0$ to $0.5$ of $\Gamma$-M line, where $k^{\prime}=k_x=k_y$. (a-e) Display the bulk energy spectrum evolution when $\mu=0.3$. The nodal superconductor can transition to second-order SC when $\Delta_\text{o}\approx0.17$ in (b), and the critical point between second-order and trivial SC phase when $\Delta_\text{o}\approx0.4$ in (d). (f-j) Exhibit the bulk energy spectrum evolution when $\mu=4$. The nodal superconductor can transition to a trivial superconductor when $\Delta_\text{o}\approx0.38$ in (i).
Common parameters: $t=1$, $\widetilde{t}=0.2$, $t^{\prime \prime}=0.1$, $\lambda_\text{I}=0.5$, $\lambda_{R}=0.5$, $\Delta_\text{i}=0.1$.}
\label{Figure3}
\end{figure*}

Starting from a nodal superconductor, the appearance of the second-order TSC or trivial superconductor can be understood in terms of the annihilation of the nodes. As we show in Fig.~\ref{Figure3} (a) and (b), the annihilation of a pair of nodes results in a phase transition from a nodal superconductor to a second-order TSC (Fig.~\ref{Figure3} (c)). The gap-closing at the $\Gamma$ point in Fig.~\ref{Figure3} (d) further leads to a trivial superconductor.
However, two pairs of nodes annihilate simultaneously corresponding to the transition from a nodal superconductor directly to a trivial superconductor, as shown in Figs.~\ref{Figure3}f-\ref{Figure3}j.

\emph{Helical TSC and second-order TSC.}---In the spirit of ``boundary of boundary'', we aim to derive an edge theory to illustrate the occurrence of corner states as Jackiw-Rebbi modes~\cite{jackiw1976solitons}. We start with an interesting observation that when $\Delta_\text{i}=0$, the system is a first-order helical TSC~\cite{meng2023soc}. The topological condition is given by $-\sqrt{\lambda_{\text{I}}^2-\Delta_\text{o}^2}<\mu<\sqrt{\lambda_{\text{I}}^2-\Delta_\text{o}^2}$. For a ${\bf k}\cdot{\bf p}$-type Hamiltonian in Eq.~\eqref{H_BdG} in the long-wave limit, we consider the edge along the $y$-axis as an example. Hence, $k_x$ can be replaced by $-i\partial_x$ while $k_y$ remains a good quantum number in the BdG Hamiltonian, $H(-i\partial_x,k_y)=H_0+H' $, where the first part 
$H_0=-i\lambda_R\partial_x\tau_z\sigma_0 s_y-\mu \tau_z\sigma_0 s_0+\lambda_{\text{I}}\tau_0\sigma_y s_z-\Delta_\text{o} \tau_y\sigma_z s_y$ needs to be solved analytically for the Majorana zero modes, and the second part $H'=-\partial_x^2(t \tau_z\sigma_0 s_0+\Delta_\text{i} \tau_x\sigma_0 s_y)-\lambda_Rk_y\tau_0\sigma_0 s_x$ is treated as a perturbation in the zero modes basis.

To solve the zero modes, a domain wall between a topologically trivial superconductor ($x<0$) and a topologically non-trivial superconductor ($x>0$) is created along the $y$-axis. For simplicity, we set the chemical potential $\mu=0$, which gives rise to a non-trivial (trivial) superconductor region with  $\Delta_\text{o}<\lambda_{\text{I}}$ 
($\Delta_\text{o}>\lambda_{\text{I}}$). Taking the ansatz $\psi(x)=\mathcal{N}e^{-\kappa x}\chi$ for the zero modes, where $\kappa_R\equiv\kappa(x>0)>0$ and $\kappa_L\equiv\kappa(x<0)<0$ since the bulk on both sides of the domain wall is gapped, and $\chi$ is the spinor part. 
After solving $H_0\psi=0$, we find only $\kappa=(\Delta_\text{o}+\lambda_{\text{I}})/\lambda_R$ satisfies the sign condition given the topological condition on both sides of the domain wall. The corresponding spinor parts are $\chi_1=(-i,0,0,-1,-i,0,0,1)^\text{T}/2$ and $\chi_2=(0,1,-i,0,0,1,i,0)^\text{T}/2$. Therefore, we arrive at the following two corresponding zero modes, $\psi_{1,2}(x)=\mathcal{N}e^{-\kappa(x)x}\chi_{1,2}$ with the normalization constant $\mathcal{N}=\sqrt{2\kappa_R\kappa_L/(\kappa_L-\kappa_R})$. 
Then, we project $H'$ onto this Majorana basis $\{\psi_1, \psi_2\}$, and find the effective edge theory
\begin{align}\label{eq-edge-ham0}
H_{\text{edge}}(k_y) = \mathcal{N}^2 \lambda_Rk_y \tau_y - m_\text{meff}\tau_x,
\end{align}
where the effective mass is $m_{\text{eff}}=-\vert\mathcal{N}\vert^2\Delta_\text{i}\delta\Delta_\text{o}/(2\lambda_R)$. 
Below, we use Pauli matrices $\tau_{x,y,z}$ for this Majorana basis. 
Please notice that the $d$-wave pairing naturally leads to a sign-changing feature for $ \delta\Delta_\text{o}\equiv\Delta_\text{o}(x<0)-\Delta_\text{o}(x>0)$ between two neighboring edges ($\Delta_\text{o}\to-\Delta_\text{o}$ under $C_{4z}$). More details can be found in the SM. Due to the sign flipping of the mass term at each corner, there will be localized zero modes at the four corners of the system, which establishes the $d+id$ second-order TSC.

\begin{figure}[!htbp]
\includegraphics[width=0.5\textwidth]{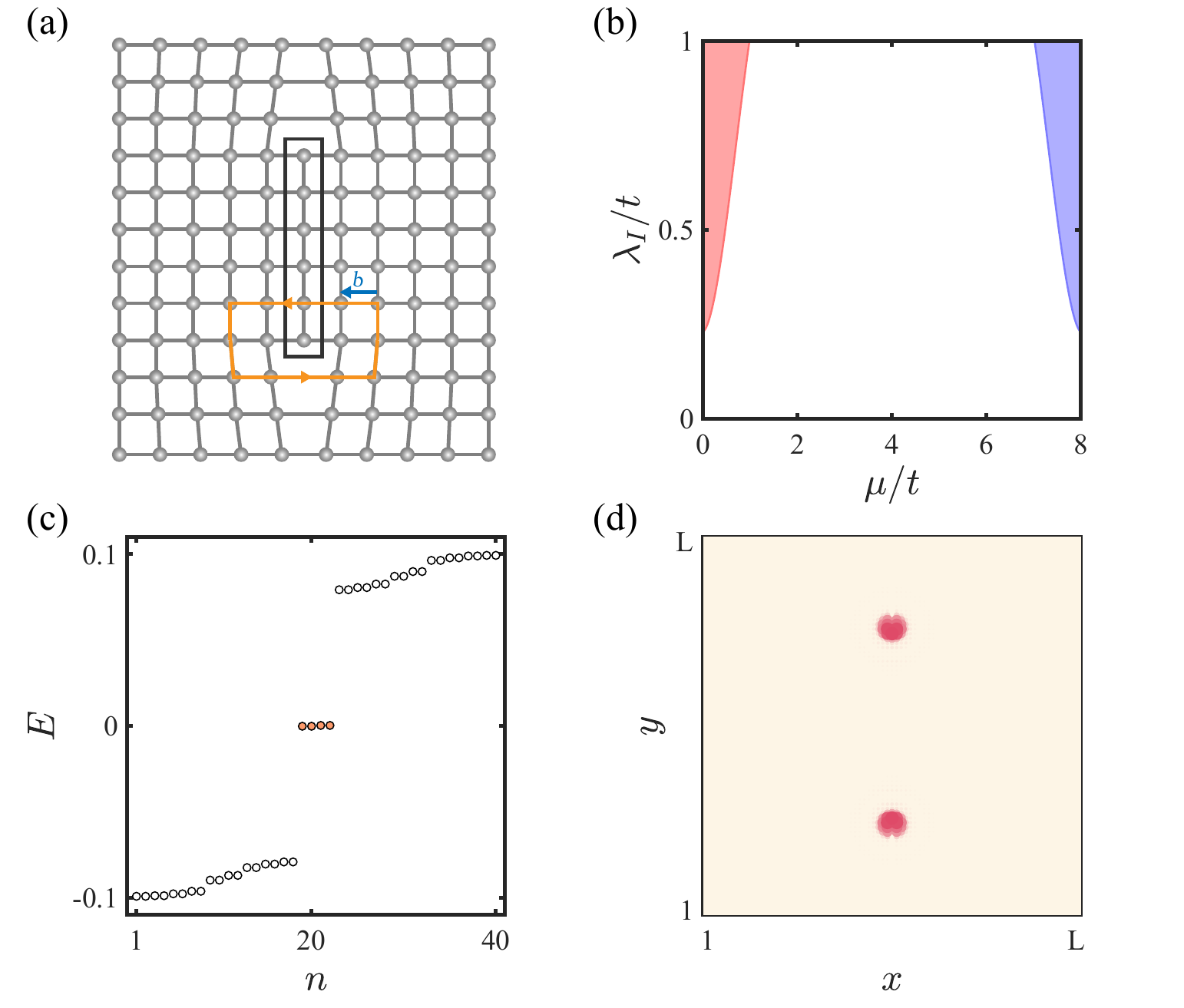}
\caption{ Topological defect on a sample of square shape. (a) The single side edge dislocation of Volterra cut-and-glue. The black orthogon displays a dislocation line. The orange and blue arrow lines show the Burgers circuit and the Burgers vector $\bm{b}=(-1,0)$. (b) Phase diagram of the trivial phase and $\mathcal{Z}_2$ non-trivial phase. The red and blue color regions are marked by different weak topological indices as $(0,0)$ or $(1,1)$ for the first-order TSC. (c) The energy spectrum of the TRS-protected first-order topological superconductor with $\Delta_\text{i}=0$. The orange dots are the fourfold dislocation Majorana zero modes. (d) The wave function profile of the four dislocation modes. Common parameters: $t=1$, $\widetilde{t}=0.2$, $t^{\prime \prime}=0.1$, $\mu=7.7$, $\lambda_\text{I}=0.5$, $\lambda_{R}=0.5$, $\Delta_\text{o}=0.3$.}
\label{Figure4}
\end{figure}

\emph{Topological defects in TSCs.}---
\label{Dislocation}
The Majorana corner modes may be hard for experimental detection, for example, the superconductivity may lose its phase coherence near the sample boundary. A bulk-defect correspondence may help to solve this issue if the bulk TSC has a non-zero weak index~\cite{Asahi2012topological,teo2013existence,hughes2014majorana,benalcazar2014classification}. For a 2D square lattice, an edge dislocation is illustrated in Fig.~\ref{Figure4} (a) with a Burger vector ${\bm{b}}=(-1,0)$. It causes an effective $\pi$-flux to trap Majorana zero modes once  
\begin{equation}\label{eq-topo-bm}
\bm{b} \cdot \bm{M_{\nu}} = 1 \textbf{ mod } 2,
\end{equation}
where $ \bm{M_{\nu}}$ is defined by $ \bm{M_{\nu}} = \nu_{1} \bm{G}_1 + \nu_{2} \bm{G}_2 $. Here, $\bm{G_{i}}$ are reciprocal lattice vectors, and the vector $(\nu_{1},\nu_{2})$ are the weak topological indices, which can be calculated via the position of Wannier center in our 2D system~\cite{PhysRevB.96.245115}. We first focus on the $\Delta_\text{o}=0$ case, the model in Eq.~\eqref{H_BdG} preserves TRS and thus belongs to class DIII of the A-Z classification~\cite{PhysRevB.55.1142}, which can be characterized by the $\mathcal{Z}_2$ topological invariant \cite{RevModPhys.88.035005}. The phase diagram in Fig.~\ref{Figure4} (b) shows $\mathcal{Z}_2=1$ for both blue and red regions, while only the helical TSC phase in the blue region ($\mu \sim 8$) carries a weak index  $(1,1)$, which is related to the polarization~\cite{nagaosa2012}. Our system has the $C_{4z}$ symmetry, thus the band inversion happens simultaneously at $(\pi,0)$ and $(0,\pi)$ points. Thus, the orbital-fluctuation-driven nematicity breaks the $C_4$ symmetry, which can be detected via this bulk-defect correspondence.
	
We then study the bulk-defect correspondence to show the dislocation Majorana zero modes. Performing a full tight-binding model calculation with an edge dislocation, the energy spectrum is shown in Fig.~\ref{Figure4} (c) and the wave function in Fig.~\ref{Figure4} (d). A periodic boundary condition for both $x$ and $y$ directions has been assumed, such that there are no Majorana corner states.
Due to the presence of TRS, Majorana Kramers pairs (MKPs) are trapped at each dislocation core. Note a dislocation with ${\bm{b}}=(0,1)$ leads to the same result.
As shown in the SM, we use the cut-and-glue progress to derive the effective 1D Hamiltonian for the MKPs. 
We first consider the ``cutting'' step.
As shown in Fig.~\ref{Figure4} (a), the edge dislocation cuts the square lattice into two parts. 
Thus, the low-energy edge Hamiltonian for the left part is given by Eq.~\eqref{eq-edge-ham0} based on the Majorana basis $\{\psi_1, \psi_2\}$. 
The mirror $M_x$ symmetry leads to that for the right part. In terms of $\{\psi_1, \psi_2, M_x \psi_2, M_x \psi_1 \}$, the low-energy Hamiltonian for this dislocation reads
\begin{align}\label{eq-dis-ham0}
H_{\text{dis}}(k_y) =  {\lambda}_{R} k_y \varrho_z \tau_y +  m_\text{eff} \varrho_0 \tau_x,
\end{align}
where $\varrho_{x,y,z}$ is for left and right part related by $M_x$. The matrix representation of $M_x$, TRS and particle-hole symmetry become $i \varrho_y \tau_x$, $i\varrho_z\tau_y \mathcal{K}$, and $\varrho_z\tau_z\mathcal{K}$ respectively.
The $m_{\text{eff}}$ term breaks TRS because of the TRS-breaking $d$-wave pairing $\Delta_\text{i}$. 
Next, we consider the effect of the ``gluing'' step on $H_{\text{dis}}$. This gives rise to the hybridization between the left edge and right edge, and $m_{\text{eff}}'(y) \varrho_x \tau_y$ is found due to the Rashba SOC [see details in the SM]. This term is allowed to preserve $M_x$ and TRS. Once Eq.~\eqref{eq-topo-bm} is satisfied, the mass $m_{\text{eff}}'(y)$ changes sign around the dislocation core~\cite{hu2022dislocation,zhang2022bulkvortex,annurevconmatphys031016025154,PhysRevB.86.100504,PhysRevB.89.224503,Chung2016,Ran2009}. When $m_{\text{eff}}=0$, $H_{\text{dis}} =  {\lambda}_{R} k_y \varrho_z \tau_y +  m_\text{eff}'(y) \varrho_x \tau_y$ corresponds to the KMP at each dislocation core.

\begin{figure}[!htbp]
\includegraphics[width=0.5\textwidth]{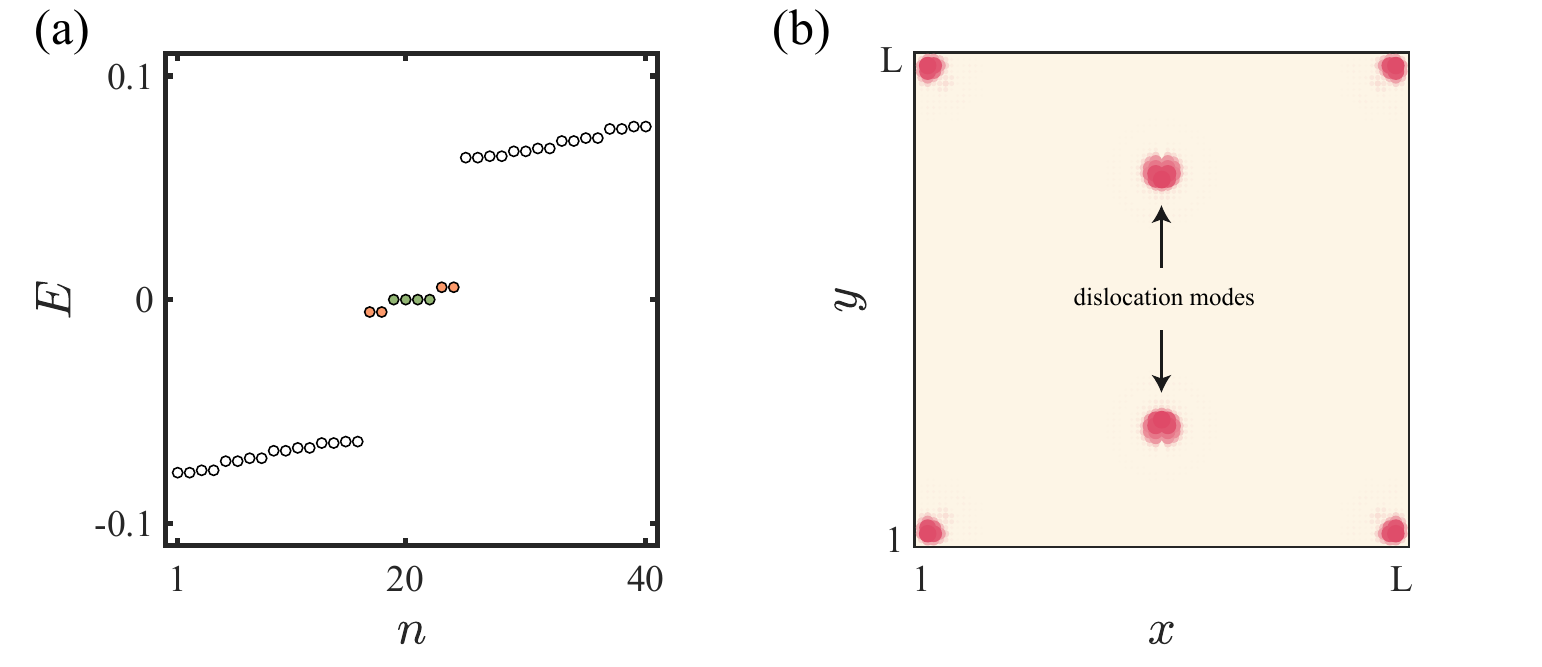}
\caption{(a) Energy spectrum of the square sample for TRS-breaking second-order TSC. The green dots energy are the fourfold degenerate zero-energy Majorana corner modes. The two pair of orange dots are gapped out dislocation modes. (b) The wave function profile of the four Majorana corner modes and four dislocation modes. Common parameters: $t=1$, $\widetilde{t}=0.2$, $t^{\prime \prime}=0.1$, $\mu=7.7$, $\lambda_{I}=0.5$, $\lambda_{R}=0.5$, $\Delta_i=0.1$ and $\Delta_o=0.3$.}
\label{Figure5}
\end{figure}

On the other hand, the dislocation KMPs will be gapped out once TRS is broken by $\Delta_\text{i} \neq 0$, as shown in Fig.~\ref{Figure5} (a). 
This gap is caused by $m_\text{eff}\propto\Delta_\text{i}$ in Eq.~\eqref{eq-dis-ham0}, since it anti-commutes with the $m_{\text{eff}}'$ term.
This is inherited from the anti-commutation relations between bulk terms in Eq.~\eqref{H_BdG}, where the Rashba term ($\lambda_R$) anti-commutes with the $d$-wave pairing ($\Delta_\text{i}$).

\emph{Conclusions.}---We have established symmetry-protected 2nd-order topology as a new topological possibility for d+id superconductors. A few candidate materials with $d$-wave pairing and spontaneous TRS breaking may be used to realize our theory, including Sr$_2$RuO$_4$~\cite{steppke2017strong}, LaPt$_3$~\cite{biswas2021chiral}, and SrPtAs~\cite{fischer2014chiral}. In the absence of TRS caused by the $d+id$ pairing, both of the two possible topological phases, the nodal TSC and the second-order TSC, can realize Majorana-bound states. We find that the nodal TSC is protected by mirror symmetry, while the second-order TSC is protected by the $C_{4z}$ symmetry, consistent with the OAI phase. The bulk-defect correspondence is also investigated using a non-zero weak index.

\emph{Acknowledgments.}---
This work was supported by the National Natural Science Foundation of China (NSFC, Grants No. 12074108,  No. 12147102, and No.~12204074), the Natural Science Foundation of Chongqing (Grant No. CSTB2022NSCQ-MSX0568), and the Fundamental Research Funds for the Central Universities (Grant No. 2023CDJXY-048). 
L.-H.~H. at Aalto is funded by the Jane and Aatos Erkko Foundation and the Keele Foundation as part of the SuperC collaboration.
R.-X.~Z. acknowledges the start-up fund at the University of Tennessee. D.-S. M. also acknowledges the funding from the China National Postdoctoral Program for Innovative Talent (Grant No. BX20220367).
	
\bibliography{bibfile}

\end{document}